\documentclass[twocolumn]{aastex63}
\usepackage[T1]{fontenc}
\usepackage[utf8]{inputenc}
\usepackage{multirow}
\usepackage{nicefrac}
\usepackage{epsfig}
\usepackage{savesym}
\savesymbol{tablenum}
\usepackage{siunitx}
\restoresymbol{SIX}{tablenum}
\usepackage{tikz}

\let\oldenddeluxetable\enddeluxetable
\let\olddeluxetable\deluxetable
\makeatletter
\renewenvironment{deluxetable}[1]{
\olddeluxetable{[#1]}
\def\pt@format{#1}%
}{\oldenddeluxetable}
\makeatother

\graphicspath{{./}{figures/}}

\begin{document}

\title{The First High-Contrast Images of X-Ray Binaries: Detection of Candidate Companions in the $\gamma$ Cas Analog RX J1744.7$-$2713}

\correspondingauthor{Myriam Prasow-\'{E}mond}
\email{myriam.prasow-emond@umontreal.ca}

\author[0000-0002-2457-3431]{M. Prasow-Émond}
\affiliation{D\'{e}partement de Physique, Universit\'{e} de Montr\'{e}al, C.P. 6128, Succ. Centre-Ville, Montr\'{e}al, QC H3C 3J7, Canada}
\affiliation{Institute for Research on Exoplanets, Université de Montréal, Département de Physique, C.P. 6128 Succ. Centre-ville, Montréal, QC H3C 3J7, Canada}

\author[0000-0001-7271-7340]{J. Hlavacek-Larrondo}
\affiliation{D\'{e}partement de Physique, Universit\'{e} de Montr\'{e}al, C.P. 6128, Succ. Centre-Ville, Montr\'{e}al, QC H3C 3J7, Canada}

\author[0000-0002-2691-2476]{K. Fogarty}
\affiliation{Division of Physics, Math, and Astronomy, California Institute of Technology, Pasadena, CA, USA}
\affiliation{NASA Ames Research Center, Moffett Field, CA 94035, USA}

\author[0000-0003-0029-0258]{J. Rameau}
\affiliation{Institute for Research on Exoplanets, Université de Montréal, Département de Physique, C.P. 6128 Succ. Centre-ville, Montréal, QC H3C 3J7, Canada}
\affiliation{Univ. Grenoble Alpes, CNRS, IPAG, F-38000 Grenoble, France}

\author[0000-0002-8323-7809]{L.-S. Guit\'{e}}
\affiliation{D\'{e}partement de Physique, Universit\'{e} de Montr\'{e}al, C.P. 6128, Succ. Centre-Ville, Montr\'{e}al, QC H3C 3J7, Canada}

\author[0000-0002-8895-4735]{D. Mawet}
\affiliation{Division of Physics, Math, and Astronomy, California Institute of Technology, Pasadena, CA, USA}
\affiliation{Jet Propulsion Laboratory, California Institute of Technology, Pasadena, CA 91109, USA}

\author[0000-0003-3105-2615]{P. Gandhi}
\affiliation{Department of Physics and Astronomy, University of Southampton, Highfield, Southampton, SO17 1BJ, United Kingdom}

\author{A. Rao}
\affiliation{Department of Physics and Astronomy, University of Southampton, Highfield, Southampton, SO17 1BJ, United Kingdom}

\author{J. F. Steiner}
\affiliation{Harvard-Smithsonian Center for Astrophysics, Cambridge, MA 02138, United States}

\author[00000-0003-3506-5667]{É. Artigau}
\affiliation{Institute for Research on Exoplanets, Université de Montréal, Département de Physique, C.P. 6128 Succ. Centre-ville, Montréal, QC H3C 3J7, Canada}

\author[0000-0002-6780-4252]{D. Lafreni\`{e}re}
\affiliation{D\'{e}partement de Physique, Universit\'{e} de Montr\'{e}al, C.P. 6128, Succ. Centre-Ville, Montr\'{e}al, QC H3C 3J7, Canada}
\affiliation{Institute for Research on Exoplanets, Université de Montréal, Département de Physique, C.P. 6128 Succ. Centre-ville, Montréal, QC H3C 3J7, Canada}

\author[0000-0002-9378-4072]{A. Fabian}
\affiliation{Institute of Astronomy, Cambridge University, Madingley Road, Cambridge CB3 0HA, United Kingdom}

\author{D. J. Walton}
\affiliation{Institute of Astronomy, Cambridge University, Madingley Road, Cambridge CB3 0HA, United Kingdom}

\author[0000-0002-3725-3058]{L. M. Weiss}
\affiliation{Department of Physics, University of Notre Dame, Notre Dame, IN 46556, USA}

\author[0000-0001-5485-4675]{R. Doyon}
\affiliation{D\'{e}partement de Physique, Universit\'{e} de Montr\'{e}al, C.P. 6128, Succ. Centre-Ville, Montr\'{e}al, QC H3C 3J7, Canada}
\affiliation{Institute for Research on Exoplanets, Université de Montréal, Département de Physique, C.P. 6128 Succ. Centre-ville, Montréal, QC H3C 3J7, Canada}

\author[0000-0003-2001-1076]{C. L. Rhea}
\affiliation{D\'{e}partement de Physique, Universit\'{e} de Montr\'{e}al, C.P. 6128, Succ. Centre-Ville, Montr\'{e}al, QC H3C 3J7, Canada}

\author[0000-0002-5837-8618]{T. B\'{e}gin}
\affiliation{D\'{e}partement de Physique, Universit\'{e} de Montr\'{e}al, C.P. 6128, Succ. Centre-Ville, Montr\'{e}al, QC H3C 3J7, Canada}

\author[0000-0002-2478-5119]{B. Vigneron}
\affiliation{D\'{e}partement de Physique, Universit\'{e} de Montr\'{e}al, C.P. 6128, Succ. Centre-Ville, Montr\'{e}al, QC H3C 3J7, Canada}

\author[0000-0003-1807-1598]{M.-E. Naud}
\affiliation{Institute for Research on Exoplanets, Université de Montréal, Département de Physique, C.P. 6128 Succ. Centre-ville, Montréal, QC H3C 3J7, Canada}

\received{ }
\revised{  }
\accepted{  }

\shorttitle{Companion Candidates in RX J1744.7$-$2713}
\shortauthors{Prasow-Émond et al.}

\submitjournal{AJ}

\begin{abstract}
X-ray binaries provide exceptional laboratories for understanding the physics of matter under the most extreme conditions. Until recently, there were few, if any, observational constraints on the circumbinary environments of X-ray binaries at $\sim$ 100--5000 AU scales; it remains unclear how the accretion onto the compact objects or the explosions giving rise to the compact objects interact with their immediate surroundings. Here, we present the first high-contrast adaptive optics images of X-ray binaries. These observations target all X-ray binaries within $\sim$ 3 kpc accessible with the Keck/NIRC2 vortex coronagraph. This paper focuses on one of the first key results from this campaign: our images reveal the presence of 21 sources potentially associated with the $\gamma$ Cassiopeiae analog high-mass X-ray binary RX J1744.7$-$2713. By conducting different analyses -- a preliminary proper motion analysis, a color-magnitude diagram and a probability of chance alignment calculation -- we found that three of these 21 sources have a high probability of being bound to the system. If confirmed, they would be in wide orbits ($\sim$ 450 AU to 2500 AU). While follow-up astrometric observations will be needed in $\sim$ 5--10 years to confirm further the bound nature of these detections, these discoveries emphasize that such observations may provide a major breakthrough in the field. In fact, they would be useful not only for our understanding of stellar multiplicity, but also for our understanding of how planets, brown dwarfs and stars can form even in the most extreme environments.

\end{abstract}


\section{Introduction} \label{sec:intro} 
Exoplanets can exist in extreme environments; in fact, \cite{1992Natur.355..145W} reported the discovery of a planetary system orbiting the millisecond pulsar PSR B1257+12. These planets were both the first confirmed exoplanets, as well as the first confirmed planets orbiting a compact object. In this system, the planets are small (from a fraction to a few $M_\oplus$) and in close orbits to the pulsar (up to a few astronomical units; hereafter AU). Currently, there are just a few ($\sim$ 7) confirmed sub-stellar companions orbiting pulsars (e.g., \citealt{2011Sci...333.1717B, 2017ARep...61..948S}). These planets are thought to originate from a fallback disk post-supernovae in which they would have formed similarly as in protoplanetary disks (e.g., \citealt{2013MNRAS.433..162Y}), or simply from gravitational capture (e.g., \citealt{2003Sci...301..193S}).

Since then, thousands of exoplanets and brown dwarfs have been discovered with the advent of new, state-of-the-art instruments. These discoveries have bettered our understanding of the Solar System and have shown that planets and brown dwarfs can exist in a variety of environments: from hot Jupiters that orbit exceedingly close to their host star (e.g., \citealt{seager_extrasolar_1998, 2018AJ....156..283E}) to planets found at orbital separations of up to several thousand AU (e.g., \citealt{lafreniere_discovery_2011, naud_discovery_2014}). Overall, these discoveries emphasize that we still do not fully understand the conditions required for the formation of planets.

A recent study argued that sub-stellar companions could even exist in extreme environments such as X-ray binaries \citep{imara_searching_2018}. These systems consist of a compact stellar remnant (white dwarf, neutron star or black hole) accreting material from a donor star, and their interaction releases strong X-ray radiation (e.g., \citealt{2006csxs.book..623T}). They have been extensively studied and monitored for several decades enabling a variety of major breakthroughs on accretion physics (e.g., \citealt{kara_corona_2019} for a recent example), as well as on the formation of outflows under extreme magnetic field conditions (see \citealt{fender_towards_2004} for a review). X-ray binaries are thus unique laboratories for studying a variety of astronomical phenomena under extreme conditions.

Although studies indicate that planets orbiting individual pulsars may be rare (e.g., \citealt{loehmer_search_2004}), a planet in close orbit to an X-ray binary could produce X-ray eclipses detectable with current instruments \citep{imara_searching_2018}. However, binary systems, such as X-ray binaries, may be more likely to host wide orbit companions (up to several hundred to thousand of AUs) because of planet-planet or planet-star interactions (e.g., \citealt{bonavita_spots_2016}). Simulations also predict that giant planets and brown dwarfs can survive stellar explosions such as those undergone by X-ray binaries, but that these may tend to push out the companions to wider orbits (e.g., \citealt{welsh_planetary_2013}). In this case, detecting sub-stellar companions via transits or radial velocity shifts may not be optimal. In addition, in the case of high-mass X-ray binaries (HMXB), the companion star has not yet exploded, implying that the system should be fairly young (i.e., less than a few dozen Myrs, e.g., \citealt{1996A&AS..117..113G}). Sub-stellar companions, if present, could therefore still be bright and amenable for detection via direct imaging (e.g., \citealt{2001RvMP...73..719B}).

As part of a pilot study aiming to explore the immediate environments of X-ray binaries, we have obtained the first high-contrast adaptive optics images of X-ray binaries ideally suited for detecting companions that have formed in situ at large ($\sim$ 100-5000 AU) radii or have been pushed out to large radii. These images consist of NIRC2 observations taken with Keck targeting all X-ray binaries within $\sim$ 3 kpc.

Here, we present the first set of observations, targeting the $\gamma$ Cassiopeiae-like HMXB harbouring a Be donor star, RX J1744.7$-$2713. In Section \ref{sec:observations}, we present the observations and data reduction. In Section \ref{sec:companions}, we show the high-contrast images of RX J1744.7$-$2713, the background/foreground contaminant probability, the color-magnitude diagram, as well as how we determined the nature of the detections. In Section \ref{sec:discussion}, we discuss our results and their implications, while in Section \ref{sec:conclusion} we summarize and discuss future perspectives.

\begin{figure*}
    \centering
    \begin{tikzpicture}
    \node(a){\includegraphics[width=\columnwidth]{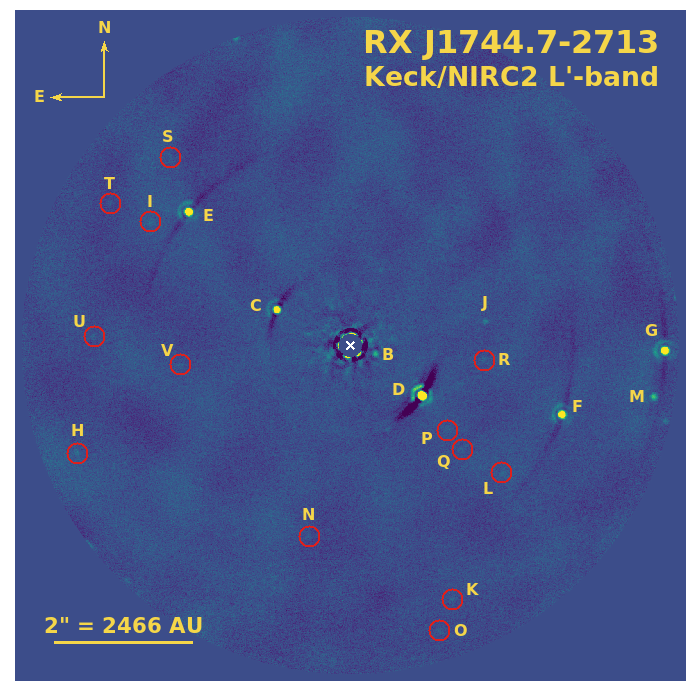}};
       
    \node at (a.south east)
    [
    anchor=center,
    xshift=-14mm,
    yshift=14mm
    ]
    {
        \includegraphics[scale=0.1]{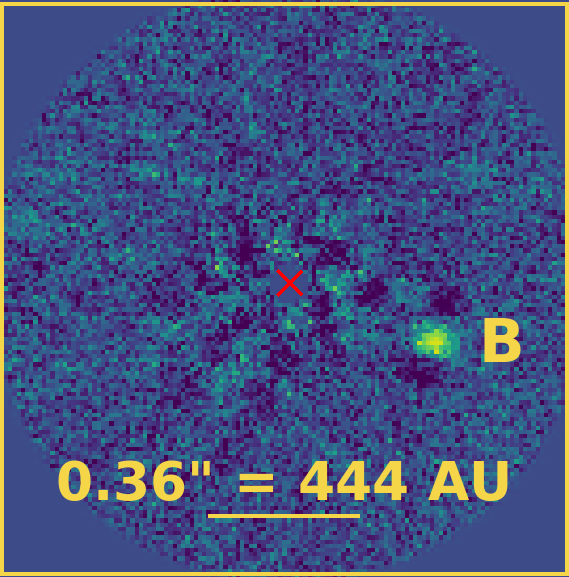}
    };
    
    \end{tikzpicture}
    \begin{tikzpicture}
    \node(a){\includegraphics[width=\columnwidth]{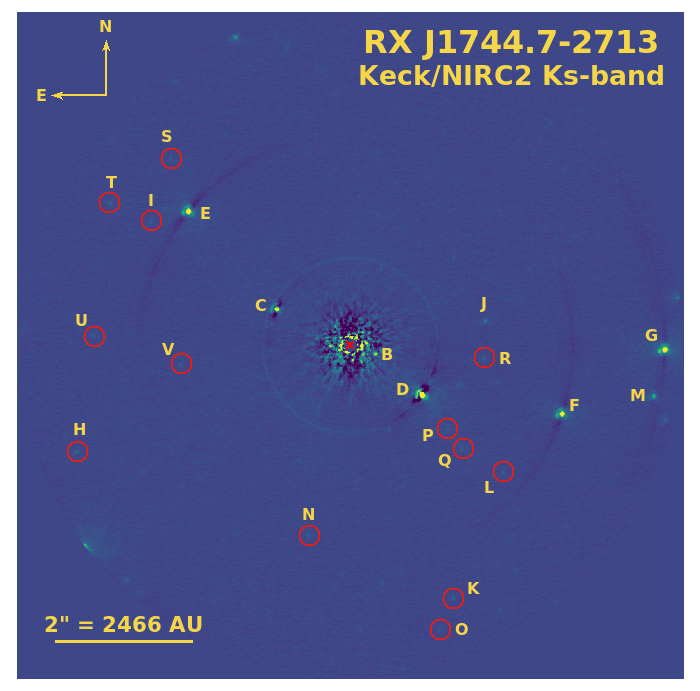}};
       
    \node at (a.south east)
    [
    anchor=center,
    xshift=-14mm,
    yshift=14mm
    ]
    {
        \includegraphics[scale=0.1]{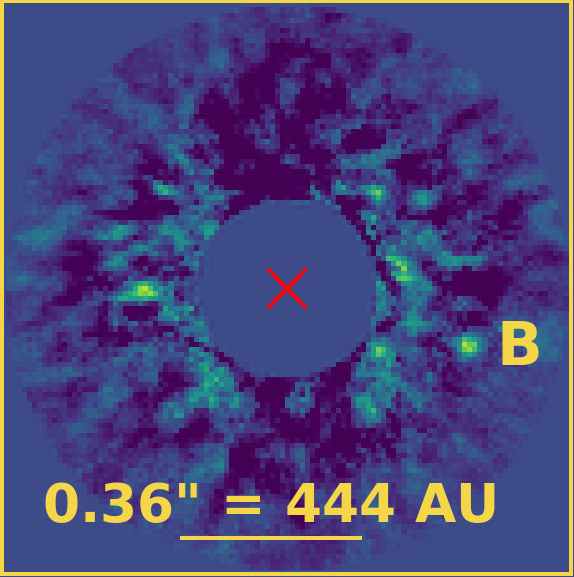}
    };
    
    \end{tikzpicture}
    
    \caption{\textit{Left}: Keck/NIRC2 $L'$ high-contrast image of RX J1744.7$-$2713 acquired on July 11 2020 and treated and reduced with a PCA annular ADI algorithm. The data were processed using \texttt{VIP} \citep{2017AJ....154....7G}. The 21 sources detected with SNR $>$ 5 when computing the signal-to-noise map are labeled. The white X symbol denotes the position of the X-ray binary masked by the coronagraph. The inset at the bottom is a zoom on the central region to focus on the closest candidate at $\sim$ 450 AU. \textit{Right}: Keck/NIRC2 $K_s$ high-contrast image of RX J1744.7$-$2713 acquired on July 12 2020. Labeled are the same objects detected from the $L'$ image. The red X symbol denotes the position of the X-ray binary masked by the coronagraph.}
    \label{fig:images}
\end{figure*}

\section{Observations and Data Reduction} \label{sec:observations}
\subsection{RX J1744.7-2713}\label{sec:RXJ1744}

RX J1744.7$-$2713 (RA: 17h 44m 45.7659s; DEC: $-$27d 13h 44.477s) is located at a distance of 1.22 $\pm$ 0.04 kpc \citep{2021A&A...649A...1G} and was first identified as a HMXB by \cite{1997ApJ...474L..53I}. The distance above is based upon geometric parallax inversion and includes a zero-point offset correction of $-$0.0348 mas according to the algorithm described in \cite{2021A&A...649A...4L}. Its optical component -- HD 161103 -- has been spectroscopically characterized in \cite{1997A&A...323..853M} and \cite{2006A&A...454..265L}, and is classified as a Be star. 

RX J1744.7$-$2713 has similar properties to $\gamma$ Cassiopeiae (also known as 2S 0053+604; e.g., \citealt{2005A&AT...24..151R, 2015ApJ...799...84S}) and has an X-ray luminosity of (3.08 $\pm$ 0.49) $\times$ 10$^{32}$ erg s$^{-1}$ \citep{naze_hot_2018}. Although the origin of X-ray emission in RX J1744.4$-$2713 is still debated, it may point to accretion onto a white dwarf (WD; \citealt{2006A&A...454..265L}). 

It should be noted that in the case of Be stars, the star's massive outflows lead to a diffuse and gaseous circumstellar disk known as a decretion disk (e.g., \citealt{2020A&A...643A.170K}), however these disks have radii at sub-AU scales, which for RX J1744.7$-$2713 is well within the inner working angle of the NIRC2 vortex coronagraph. 

In the case of RX J1744.4$-$2713, \cite{coleiro_distribution_2013} estimated the age of the system to be $\sim60$ Myr by investigating the expected offset between the position of the Galaxy's spiral arms and HMXBs. However, considering the uncertainties (e.g., see their Fig. 13), the age of RX J1744.4$-$2713 is likely to be anywhere between a couple Myrs to 80 Myr (95\% confidence). Given these uncertainties, we consider the range of 5--60 Myr throughout this study and note that even if we were to consider an age outside this range, our conclusions remain the same.

Furthermore, the system has proper motions of $-0.955 \pm 0.100$ mas/yr and $-2.062 \pm 0.077$ mas/yr in RA and DEC respectively and is located in the Galactic plane, as most of the HMXBs in our galaxy (e.g., \citealt{2006csxs.book..623T}).

\subsection{Observations}\label{sec:obs}
The Keck/NIRC2 vortex coronagraph \citep{2005ApJ...633.1191M, 2016OptCo.379...64S} was used to observe RX J1744.7$-$2713 in pupil-tracking mode -- with the narrow-field camera (9.971 $\pm$ 0.004 mas/pixel; \citealt{2016PASP..128i5004S}) -- in $L'$-band in 2017 ($\lambda$ = 3.776 $\mu$m, $\Delta \lambda$ = 0.700 $\mu$m; PI: Mawet) and in both $L'$-band and $K_s$-band in 2020 ($\lambda$ = 2.146 $\mu$m, $\Delta \lambda$ = 0.311 $\mu$m; PI: Fogarty). Observations were adaptive-optics (AO)-assisted with the Keck II Shack-Hartmann wavefront sensor in 2017, which performs wavefront sensing in $R$, and the infrared pyramid wavefront sensor (PyWFS) in 2020, which performs wavefront sensing in $H$ \citep{2000PASP..112..315W, 2018SPIE10703E..1ZB}. Given the $R$ and $H$ magnitudes of RX J1744.7$-$2713 ($R \sim 9$; \citealt{2003yCat.2246....0C}, $H \sim 7$; \citealt{2012yCat.1322....0Z}), the performances of the two wavefront sensors are comparable; however, we chose to use the PyWFS for our 2020 observations since sensing in $H$ is advantageous for several extremely red HMXBs in our 2020 survey sample. Tip-tilt adjustments were performed with Quadrant Analysis of Coronagraphic Images for Tip-tilt sensing (QACITS; \citealt{2017A&A...600A..46H}) in order to keep the star well-centered on the vortex focal plane mask.

Individual exposure times, coadds, total number of frames and total integration times obtained for each observation are summarized in Table \ref{tab:observations}. 

In our 2017 observations, we only obtained $\sim 14^{\circ}$ of field rotation, but in our 2020 observations we obtained $\sim 40^{\circ}$ of rotation in both bands to enable reduction using angular differential imaging (ADI; \citealt{2006ApJ...641..556M}). Additional frames taken with RX J1744.7$-$2713 moved off-axis were taken throughout each run to sample the evolution of the point-spread function (PSF) throughout the run and to obtain photometric measurements of RX J1744.7$-$2713. 

\begin{table}[h!]
\caption{Keck/NIRC2 Observing Log for RX J1744.7$-$2713. Abbreviations: Wave Front Sensor (WFS), Parallactic angle coverage (P. A. cov.)}
\centering
\begin{tabular}{cccc}
\hline\hline
UT Date & 2017 Sep 8 & 2020 Jul 11 & 2020 Jul 12\\ \hline
Filter & $L'$ & $L'$ & $K_s$\\
WFS & SH & py & py \\
Int. Time (s) & 0.5 & 0.5 & 0.6 \\
Coadds & 60 & 60 & 45 \\
Total frames & 40 & 120 & 92\\
P. A. cov. ($^\mathrm{o}$) & 14.3 & 38.3 & 39.6 \\ \hline 
\end{tabular}
\label{tab:observations}
\end{table}

\newpage
\subsection{Data Reduction}\label{sec:datareduction}
Data reduction (flat-fielding, bad pixel masking, sky removal, 
image registration and selection, and centering via speckles) was performed using the Vortex Image Processing (\texttt{VIP}) package \citep{2017AJ....154....7G}. The frames were then de-rotated and mean-combined, and a Principal Component Analysis (PCA) annular ADI algorithm from \texttt{VIP} was applied to obtain the high-contrast images (see Figure \ref{fig:images}). Note that the number of principal components ($n_\mathrm{comp}$) for each individual source was chosen to reach an optimal signal-to-noise ratio (SNR), by testing many values ranging from 1 to 50. We then generated the SNR and significance (signal in terms of $\sigma$; \citealt{2014ApJ...792...97M}) maps to identify sources with a SNR greater than 5 (which results in a signal greater than 4$\sigma$ for all of our sources). This also provided estimates of the sources' coordinates. 

\subsection{Magnitude Calculation}\label{sec:mag}
To calculate the full width at half maximum (FWHM) of the PSF for both filters, we fit a gaussian profile to the median-collapsed PSF data cube. We obtained a FWHM of 83.2 milliarcseconds (hereafter mas) and 47.5 mas for $L'$ and $K_s$ respectively. Note that we also generated a 2D synthetic PSF for $L'$ flux fitting -- using a gaussian model and the corresponding FWHM -- because of an artefact caused by thermal background noise from the primary mirror that has been folded back on-axis by the vortex phase mask. However, the use of a synthetic PSF does not affect the flux fitting much, as we get consistent results in the error bars when we test our $K_s$ data with both real and synthetic PSFs.  

Using the reduced data cube and the normalized and centered synthetic PSF, we injected fake companions with an inverse flux into the science data cube to estimate the photometric counts, coordinates, and errors \citep{2010Sci...329...57L} of any detected source with a SNR $>5$ in the $L'$ SNR map. This process uses a Nelder-Mead optimization algorithm to find the coordinates and the flux of a source in a given aperture (a radius of 3 FWHM). We then divided the counts of the sources by the counts of the HMXB's median-collapsed PSF data cube -- measured in the same aperture as the sources -- in order to obtain the flux of the candidate companion relative to the X-ray binary (i.e., the contrast).

Knowing the apparent magnitudes of RX J1744.7$-$2713 ($L'$ = 5.809 $\pm$ 0.077 mag and $K_s$ = 6.507 $\pm$ 0.023 mag; \citealt{2003yCat.2246....0C}), the contrast can be converted to apparent magnitudes and absolute magnitudes (using the known distance in parsec), and then to mass using a model from MESA Isochrones \& Stellar Tracks (MIST; \citealt{2011ApJS..192....3P, 2013ApJS..208....4P,2018ApJS..234...34P, 2015ApJS..220...15P, 2016ApJS..222....8D, 2016ApJS..225...10F, 2016ApJ...823..102C}).

\cite{2012int..workE..23S} reported that there is no significant variability in the light curve of HD 161103 (Be star in RX J1744.7$-$2713). However, large time scale ($\sim$ 50--100 yr) variability could still occur. We therefore injected a 5\% additional error to the magnitudes to occur for any large time scale variability, and adding even larger errors does not change our results. As for the line-of-sight interstellar reddening $E(B-V)$, \cite{2011ApJ...737..103S} estimated it to be $E(B-V) \sim 3.4$. Since the extinction $A(V)$ is related to $E(B-V)$ via $A(V) = 3.2 E(B-V)$, we have $A(V) \sim 10.9$, and assuming a typical extinction power law varying as $A(V) \propto \lambda^{-1.75}$ (\citealt{1989ESASP.290...93D}; and see \citealt{1994MNRAS.266..497D, 2005A&A...433..117V, 2017MNRAS.471.3617D, 2020AJ....160..283U} for examples of papers that use that technique), we obtain $A(K_s) \sim 0.85$ and $A(L’) \sim 0.32$ (via a simple cross product). Therefore, $K_s$ and $L'$ are not strongly impacted by dust reddening; such an extinction results in a color shift of $\sim 0.5$ mag for $K_s - L'$ and a magnitude shift of $\sim$ 0.3 for $L'$. These corrections have been applied to our results; by adding them to the $K_s$ and $L'$ magnitudes.

\subsection{Injection/Recovery}\label{sec:injection}

In order to estimate the errors on the parameters -- the flux and the position (i.e., the separation from the XRB in mas and the parallactic angle in degrees) -- obtained with the optimization process, we proceeded by injection recovery. We first injected fake companions with known values for the parameters, and then ran the optimization code to compare the outputs with the known inputs. For each parameter, we tested this method for 90 different values, with a range covering the values of the true candidate companions of the system. We found that the errors were relatively stable while increasing the values of the parameters, meaning they are more significant for smaller values within the range. Averaging the results, we obtained errors of the order of $\sim$ 0.5 mas for the separation and $\sim$ 0.01$^\mathrm{o}$ for the position angle. As for the flux, we considered only the errors explained in Section \ref{sec:mag}, as they are dominant and physically significant.

The injection/recovery method gives relatively small errors, but other types of error must be included (e.g., the error when determining the center of the vortex coronagraph within the different frames). As our 2020 $L'$ and $Ks$ observations are spaced out by only one day, the astrometry parameters should be the same. Therefore, we also considered the difference between our 2020 $L'$ and $Ks$ results as part of astrometry errors, which are greater than the injection/recovery errors.

\section{Companion candidates}\label{sec:companions}
\subsection{High-Contrast Images} \label{sec:images}
The left panel of Figure \ref{fig:images} presents the $L'$ high-contrast image of RX J1744.7$-$2713, while the right panel shows the $K_s$ image, both from the 2020 observation runs. The $L'$ image from 2017 is shown in the Appendix, in which only 3 sources are detected. With deeper observations in 2020 (unlike the poor parallactic coverage for angular differential imaging in 2017), it revealed a then-unresolved from the central X-ray binary $\sim$ 450 AU candidate companion. We therefore only consider the 2020 observations for the remainer of the paper, except for a preliminary proper motion analysis in Section \ref{sec:pma}. 

Using the $L'$ SNR map, we find a total of 21 sources that have a SNR $>$ 5. These sources are labeled from B to V in Fig. \ref{fig:images} and are also detected in the $K_s$ image with a SNR $>$ 5. Moreover, as the $K_s$ image is more sensitive (more sources), note that we only considered the sources detected in $L'$ when measuring the magnitudes, since potential sub-stellar companions are usually brighter in $L'$ \citep{2016ApJS..225...10F}. The projected separations from RX J1744.7$-$2713, estimated using the distance and the plate scale, range from $\sim$ 450 to $\sim$ 5500 AU.

In the next sections, we examine the nature of these sources via different analyses.

\subsection{Background Source Contamination}
\subsubsection{TRILEGAL}\label{sec:TRILEGAL}
To assess the possibility that the field of view may be dominated by background or foreground sources, we derived an estimation of the expected number of sources for the field of view using TRILEGAL -- a 3D model used for simulating the stellar photometry of any Galaxy field (\citealt{2005A&A...436..895G}; and see \citealt{2015A&A...578A..51C, 2018A&A...620A.102D, 2021MNRAS.504..565J, 2021ApJS..253...53W} for examples of papers that use this model). Constraining the model to the faintest detected magnitude in $L'$, we calculated the expected number of sources for the entire field of view (i.e. 10.24'' $\times$ 10.24''). Fig. \ref{fig:TRILEGAL} presents the distribution of the $L'$ magnitude for the simulated sources and for RX J1744.7$-$2713. 

We first conducted a Kolmogorov-Smirnov test to assess the possibility that the two distributions follow the same statistical function. Regardless of the horizontal binning size, we obtain a very small probability ($\ll 0.1)$ of observing the same distribution. Therefore, we can reject the null hypothesis and conclude that at least one of the sources is likely to be bound. However, we note that for fainter magnitudes, the distributions are roughly similar, so the corresponding sources are likely unrelated sources. On the contrary, there is a larger proportion of brighter objects than predicted by TRILEGAL. We thus conclude that the brighter sources in RX J1744.7$-$2713 (C, D, E, F, and G in Fig. \ref{fig:images}) are less likely to be contaminants, as their presence is not explained by the model. However, this method is not precise and is only intended to provide a first rough estimate.

\begin{figure}[ht]
    \centering
    \includegraphics[width = \columnwidth]{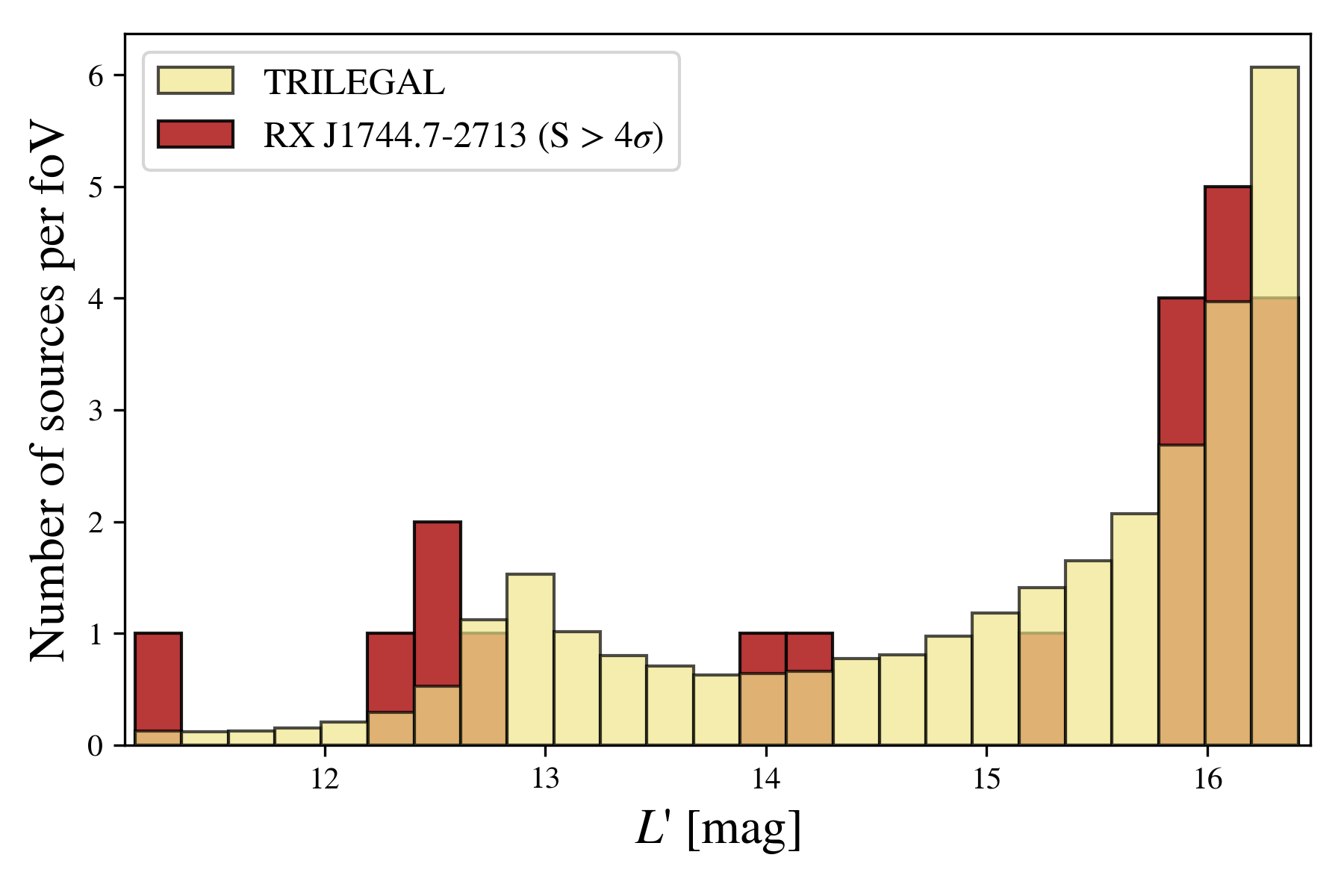}
    \caption{$L'$ distribution for RX J1744.7$-$2713 (red) and for the sources simulated by TRILEGAL (yellow). The orange color is where the two distributions are superposed.}
    \label{fig:TRILEGAL}
\end{figure}

\subsubsection{Probability of Chance Alignment}
In addition to TRILEGAL, we derived a probability of chance alignment -- i.e., the likelihood that the candidates are not bound to the system -- using the Two Micron All-Sky Survey (2MASS) Point Source Catalog (PSC; see \citealt{2006A&A...459..909C, 2008ApJ...683..844L, 2014ApJ...785...47L} for examples of that technique). More specifically, we retrieved all the sources in a 15' $\times$ 15' field surrounding RX J1744.7$-$2713, and we constructed a cumulative distribution of the number of sources as a function of the apparent $K_s$ magnitude. We then divided this result by the area (in arcsec$^2$) to obtain a surface density $\Sigma$ that depends on the limiting magnitude of the candidate, i.e., $\Sigma$ = $\Sigma(K_s < Ks_\mathrm{candidate})$. Assuming a random uniform Poisson distribution of unrelated objects across the chosen area, the probability that a candidate is not related to the system is given by:

\begin{equation}
    P_\mathrm{unrelated}(\Sigma, \Theta) = 1 - \exp(-\pi\Sigma\Theta^2)
    \label{eq:chance}
\end{equation}

where $\Theta$ is the projected separation from RX J1744.7$-$2713 (in arcsec). Table \ref{tab:AC} shows the $\Sigma$, $\Theta$, and $1 - P_\mathrm{unrelated}(\Sigma, \Theta)$ values for all sources (B to V).

\begin{table}[ht!]
\centering
\caption{Probability of unrelation with RX J1744.7$-$2713 for all the sources (B to V in Fig. \ref{fig:images}), as estimated from 2MASS PSC.}
\begin{tabular}{cccc}
\hline\hline
Source & $\Sigma$ & $\Theta$ & $1 - P_\mathrm{unrelated}(\Sigma, \Theta)$ \\ 
& (arcsec$^{-2}$) & (arcsec) & (\%)\\ \hline
B & 3.24 $\times$ 10$^{-2}$ & 0.37 & 98.6 \\
C & 2.11 $\times$ 10$^{-2}$ & 1.18 & 91.2 \\
D & 7.14 $\times$ 10$^{-3}$ & 1.25 & 96.6 \\
E & 1.66 $\times$ 10$^{-2}$ & 3.02 & 62.3 \\
F & 2.47 $\times$ 10$^{-2}$ & 3.19 & 45.5 \\
G & 2.43 $\times$ 10$^{-2}$ & 4.51 & 21.2 \\
H & 3.29 $\times$ 10$^{-2}$ & 4.23 & 15.8 \\
I & 3.29 $\times$ 10$^{-2}$ & 3.37 & 30.9 \\
J & 3.29 $\times$ 10$^{-2}$ & 1.96 & 67.3 \\
K & 3.29 $\times$ 10$^{-2}$ & 3.93 & 20.2 \\
L & 3.29 $\times$ 10$^{-2}$ & 2.85 & 43.2 \\
M & 3.29 $\times$ 10$^{-2}$ & 4.42 & 13.4 \\
N & 3.29 $\times$ 10$^{-2}$ & 2.80 & 44.4 \\
O & 3.29 $\times$ 10$^{-2}$ & 4.29 & 14.9 \\
P & 3.29 $\times$ 10$^{-2}$ & 1.83 & 70.6 \\
Q & 3.29 $\times$ 10$^{-2}$ & 2.18 & 61.3 \\
R & 3.29 $\times$ 10$^{-2}$ & 1.91 & 68.6 \\
S & 3.29 $\times$ 10$^{-2}$ & 3.74 & 23.6 \\
T & 3.29 $\times$ 10$^{-2}$ & 4.02 & 18.9 \\
U & 3.29 $\times$ 10$^{-2}$ & 3.69 & 24.5 \\
V & 3.29 $\times$ 10$^{-2}$ & 2.48 & 53.1 \\ \hline
\end{tabular}

\label{tab:AC}
\end{table}

This table shows that sources B, C, D have a high probability of not being unrelated to the system, respectively of 98.6\%, 91.2\%, and 96.6\%, which is consistent with Fig. \ref{fig:TRILEGAL}. Moreover, sources E, J, P, Q, and R have high but less convincing probabilities, while the remaining sources are likely to be background stars since their probability is lower than 60\%. 

\subsection{Color-Magnitude Diagram}\label{sec:diag}

\begin{figure}[ht!]
    \centering
    \begin{tikzpicture}
    \node(a){\includegraphics[width=\columnwidth]{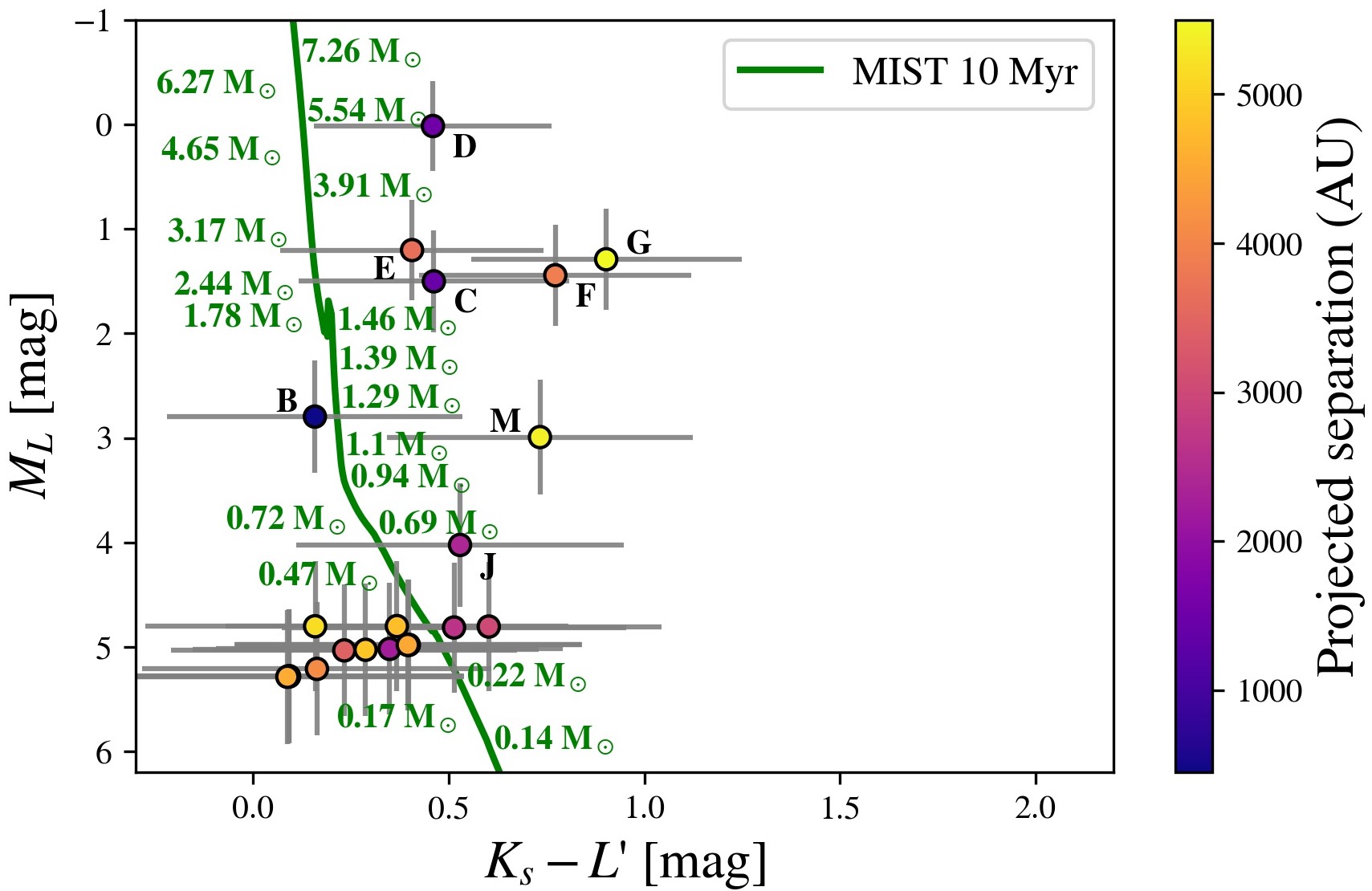}};
       
    \node at (a.south east)
    [
    anchor=center,
    xshift=-30mm,
    yshift=39mm
    ]
    {
        \includegraphics[scale=0.2]{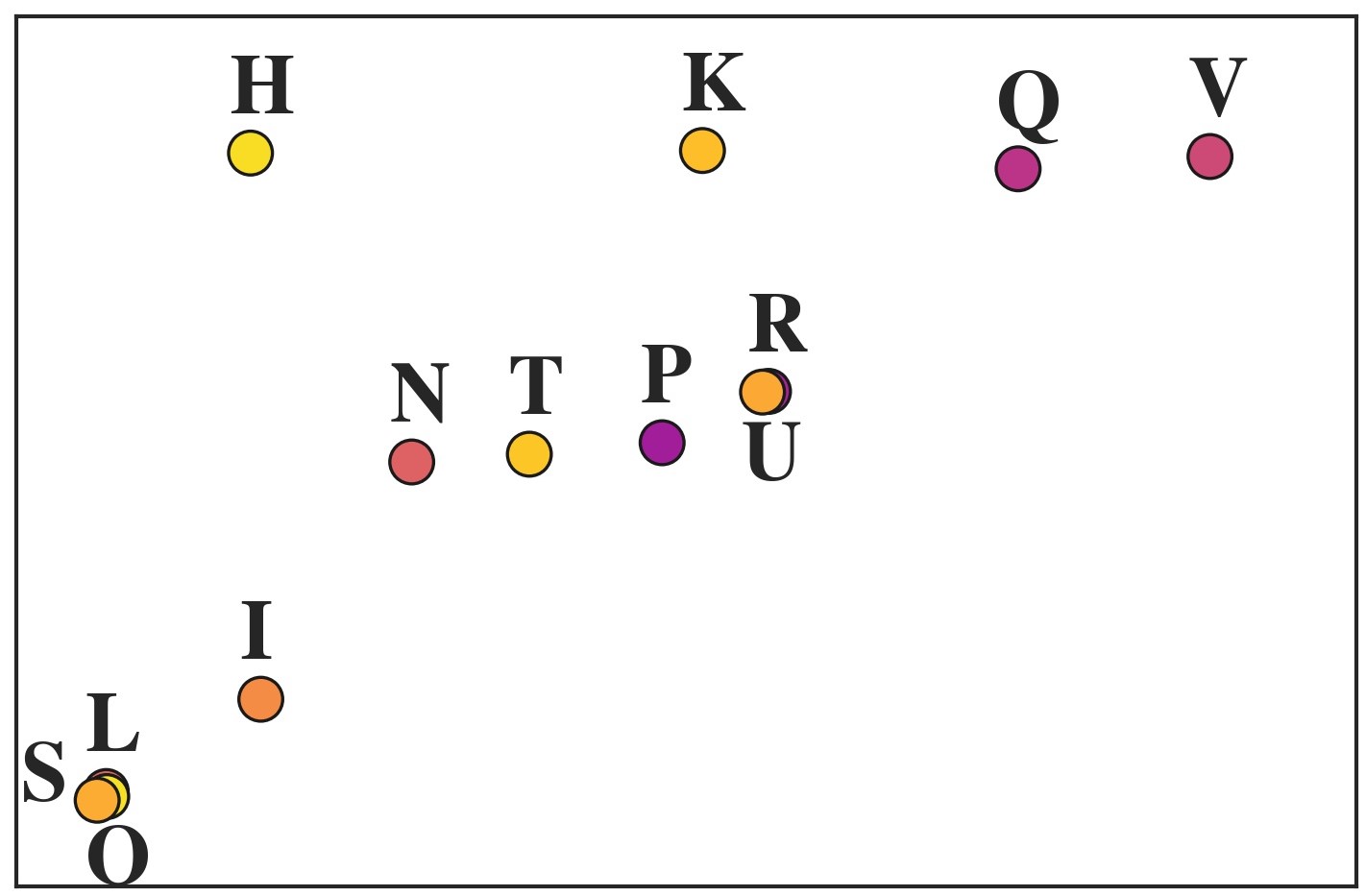}

    };
    
    \node at (a.south east)
    [
    anchor=center,
    xshift=-30mm,
    yshift=22mm
    ]
    {
        \includegraphics[scale=0.2]{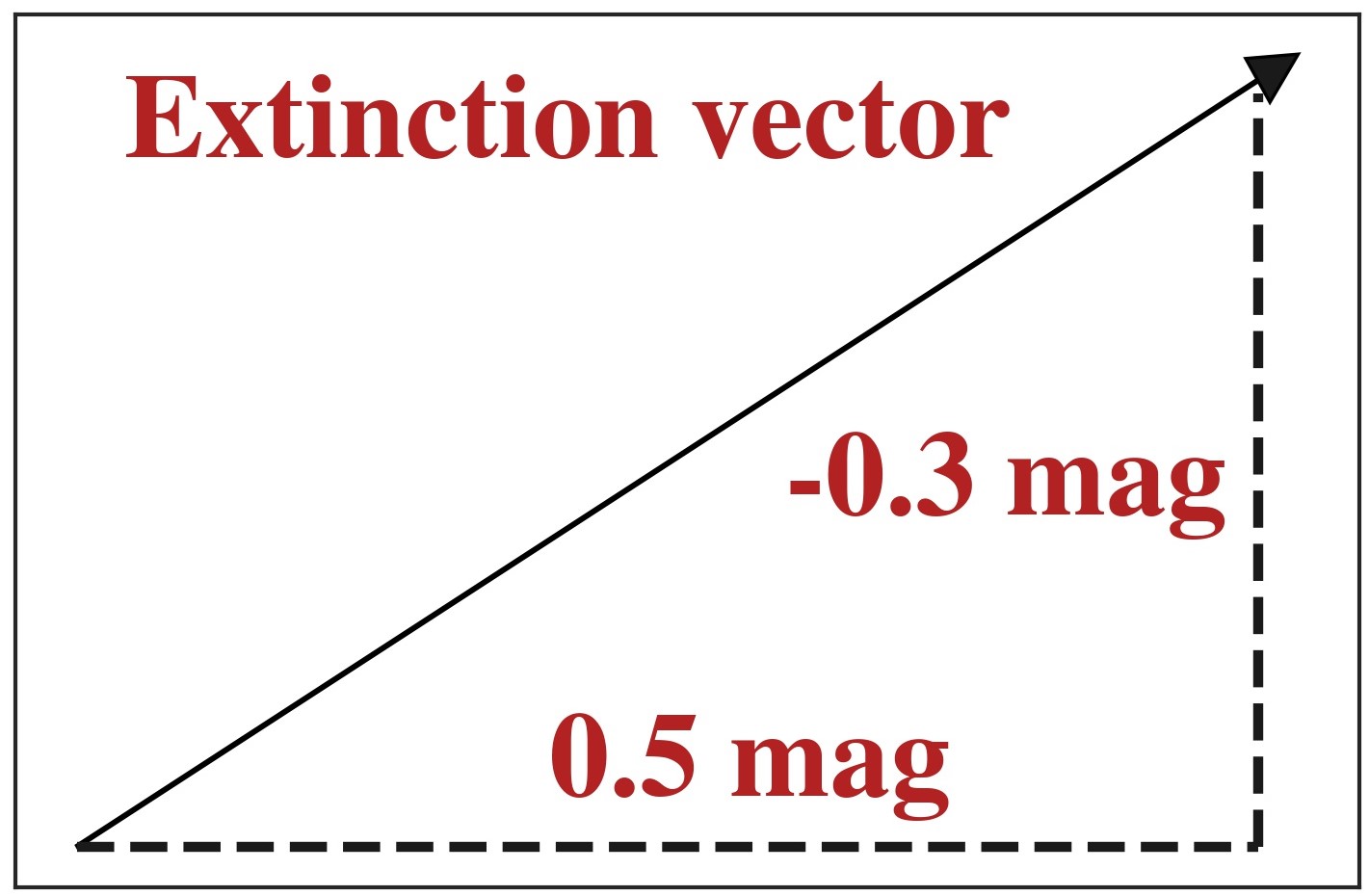}

    };
    
    \end{tikzpicture}
    \caption{Absolute magnitude in $L'$ versus $K_s-L'$ color for the 21 companion candidates (colored circles) in RX J1744.7$-$2713. The colorbar indicates the projected separation (in AU) from the X-ray binary. Overlaying is the MIST evolutionary track for stars at 10 Myr \citep{2011ApJS..192....3P, 2013ApJS..208....4P,2018ApJS..234...34P, 2015ApJS..220...15P, 2016ApJS..222....8D, 2016ApJS..225...10F, 2016ApJ...823..102C}. The inset at the top-right is a close-up of the fainter sources.} The inset at the bottom-right shows the extinction vector, i.e. the shift in magnitude and color caused by extinction (corrections have already been applied).
    \label{fig:colormag}
\end{figure}

Since both $L'$ and $K_s$ observations were taken, we were able to construct a color-magnitude diagram, as has been done in previous studies to estimate the physical properties of companions (e.g., \citealt{2013ApJ...772L..15R, 2018MNRAS.480.3170S, 2021MNRAS.504..565J}). Figure \ref{fig:colormag} presents the absolute magnitude in $L'$ versus $K_s-L'$ for the 21 potential companions to RX J1744.7$-$2713. The track from MIST at 10 Myr is overlaid, as well as the extinction vector (see Section \ref{sec:mag}).

From this diagram, the fainter sources (H to L and O to V) could be coherent with low-mass stars (M- or K-type), while B and M could be G-type stars. Moreover, B, C, E, F, and G have estimated masses potentially coherent with massive A- or B-type stars. Due to their off-set $K_s - L'$ color, the other sources (F, G, M) seem to be inconsistent with being sub-stellar or stellar companions, and may therefore be unrelated background objects such as extincted background stars or artefacts. 

A Table with the position, the $L'$ and $K_s$ magnitudes, the estimated mass, and the projected separation of the 21 sources companion candidates is presented in the Appendix. Note that the masses were estimated from the 5--60 Myr stated in Section \ref{sec:RXJ1744} and that we did not apply the extinction correction for this Table. 

Thus, RX J1744.7$-$2713 does not currently show evidence of exoplanet-mass companions, as the candidate companions have masses from $\sim$ 0.2 to 6 $M_\mathrm{\odot}$. However, we reached excellent SNR and we detected sources down to $L' \sim 16.5$. Another way to gather more information on the sources is via additional bands (e.g., $H$) or via spectroscopic characterization.

\subsection{Preliminary Proper Motion Analysis}\label{sec:pma}
To confirm more robustly that the sources are bound, a proper motion analysis could be conducted (e.g., \citealt{2008Sci...322.1348M}). Given the distance ($\sim 1$ kpc) and proper motion of RX J1744.7$-$2713 (see Section \ref{sec:RXJ1744}), a $>$ 3$\sigma$ detection will only be possible in $\sim$ 5--10 years. However, considering that there is a three-year interval between our 2017 and 2020 sets of observations, a preliminary proper motion analysis was conducted for sources C, D, and J. The plots of astrometric positions (position angle in degrees, separation in mas, and relative separations in right ascension and declination) for each of these sources relative to the X-ray binary can be found in the Appendix. Accounting for the errors from Section \ref{sec:injection}, we conclude that both scenarios -- stationary background or comoving with RX J1744.7$-$2713 -- are possible for sources D and J at this current epoch. We note that the position of source C is offset from the tracks, which indicates that it could be a non-stationary background object.

\subsection{Stability of the Wide Orbits}
The 21 sources detected in RX J1744.7$-$2713 at high SNR would have wide projected separations (from $\sim$ 450 AU to $\sim$ 5500 AU) which raises concerns as to whether potential companions could be gravitationally bound to the system. However, very large separations have been observed in many known systems including companions white dwarfs or other evolved systems (e.g., 2000 AU for GU Psc b; \citealt{naud_discovery_2014}, 2985 AU for WISE 2005+5424; \citealt{2013ApJ...777...36M}, 16 500–-26 500 AU for LSPM 1459+0857 AB; \citealt{2011MNRAS.410..705D}). \cite{2019AJ....158..187B} reported that the occurrence of sub-stellar companions at very large orbits are rare. However, in RX J1744.7$-$2713 the mass ratios are small (from $q \sim 0.003$ to $\sim$ 0.08), and such ratios have yet to be studied in detail. 
Despite the large separations, the large mass of the central system (i.e., the X-ray binary) implies a high binding energy. Assuming a total mass $M_\mathrm{XRB}$ of $\sim$ 18 $M_\odot$ for RX J1744.7$-$2713 (typical masses for WD and Be stars) and a mass $M_c$ of $\sim$ 80 $M_\mathrm{Jup}$ as a lower limit for the companions, we can estimate the binding energy of the system via the following equation (assuming circular orbits):

\begin{equation}
    E_\mathrm{bind} \sim -\frac{GM_\mathrm{XRB}M_c}{1.27r}
\end{equation}

Here, 1.27 is the average projection factor between $r$ and the semi-major axis, assuming a random viewing angle (e.g., \citealt{2006ApJ...652.1572B}). We obtain $E_\mathrm{bind}$ $\sim $ $-$4.3 $\times$ 10$^{43}$ ergs and $E_\mathrm{bind}$ $\sim $ $-$3.2 $\times$ 10$^{42}$ ergs for the closest (450 AU) and farthest (5500 AU) candidates respectively. In both cases, the binding energy is higher than in many known bound systems (e.g., $\sim 10^{41}$ erg;  \citealt{naud_discovery_2014}).

Therefore, it is not unreasonable that the multiple stellar companions candidates, if confirmed with proper motion analysis, are gravitationally bound to the HMXB despite their wide separations. However, it is important to note that it is unlikely that all these sources would be bound due to dynamic instability. In fact, although we do not have detailed information on their potential orbital motion yet, candidate companions C and D have similar separations from RX J1744.7$-$2713, and simulations show that this configuration can be unstable (e.g., \citealt{1994MNRAS.267..161K}). On the contrary, a scenario of a triple system -- composed of the X-ray binary (considered as one system here), candidate companion B, and candidate companion C, D, E or J -- is more plausible according to basic $N$-body simulations.

\section{Discussion} \label{sec:discussion}

These analyses allowed us to reject most of the 21 sources as candidate companions. Until a more significant proper motion analysis can be undertaken, we consider sources B, D and J to be strong candidates of stellar companions bound to RX J1744.7$-$2713 and we discuss the implications of these results in the following section. 

\subsection{Stellar Multiplicity}
The discovery of multiple probable stellar candidate companions in RX J1744.7$-$2713 implies that it would be a multiple-star system. In fact, high order multiplicity increases with primary mass (e.g., \citealt{2012A&A...538A..74P}), and although surveys are incomplete, they suggest that stellar multiplicity is frequent in high-mass stars such as young B stars (see \citealt{2013ARA&A..51..269D} for a review on stellar multiplicity). More precisely, if we consider only the confirmed companions, the multiplicity frequency would be $\geq$ 80\% for stars with a mass $\geq$ 16 $M_\mathrm{\odot}$ (e.g., \citealt{2012MNRAS.424.1925C}). Regarding the companion frequency (CF; i.e., the average number of companion per target in a population), it is $\approx$ 130 $\pm$ 20\% for high-mass stars \citep{2013ARA&A..51..269D}. Moreover, the frequency $N(n)$ of multiplicity $n$ roughly follows a geometric distribution $N(n) \propto 2.3^{-n}$ (up to $n = 7$; \citealt{2008MNRAS.389..869E}) for solar-type stars; however there is currently no derived relation for massive stars. As RX J1744.7$-$2713 is already a known high-mass binary system, the complete sample of this project could therefore enable us to better understand if X-ray binaries are likely to stellar multiplicity, as well as probing stellar multiplicity at mass ratios below $\sim$ 0.1.

\subsection{Similarities to $\gamma$ Cassiopeiae}
In addition, RX J1744.7$-$2713 is a $\gamma$ Cassiopeiae analog, and the latter has two faint optical companions (e.g., \citealt{2001AJ....122.3466M}). Therefore, it suggests that Be stars might be more favorable to stellar multiplicity. As $\gamma$ Cassiopeiae is a HMXB candidate (e.g., \citealt{1995A&A...296..685H}), it also means that stellar companions could be common in X-ray binaries.

\subsection{Formation Scenarios}

In the case of X-ray binaries, companions are believed to form (1) from the direct fragmentation of a collapsing pre-stellar core, i.e., roughly at the same time as the stars that will eventually form the X-ray binary (e.g., \citealt{2012MNRAS.419.3115B}), (2) from the circumbinary disk surrounding the initial binary system (i.e., pre-explosion of the then-compact object; e.g., \citealt{2010ApJ...708.1585K}), or (3) from the supernova fallback disk surrounding the X-ray binary (post-explosion; e.g., \citealt{1992Natur.355..145W}). Note that fallback disks could lead to the formation of sub-stellar companions, but they are not massive enough to allow star formation. However, this work does not show evidence of such disks, so we can not conclude how the candidate companions have formed. Furthermore, if companions are formed under the last scenario, then they must be younger than the X-ray binary; their age could be lower than 5 Myr (the lower limit for RX J1744.7$-$2713's age stated in Section \ref{sec:RXJ1744}).

Overall, X-ray binaries could host multi-generation companions that form at different epochs through the evolution of the binary. This work explores the idea that the interactions between X-ray binaries and their surroundings might be complex, leading to multiple companions and hierarchical orbits. Moreover, it opens up the discussion on how sub-stellar and stellar formation could occur even in the most extreme environments.

\section{Summary and conclusions} \label{sec:conclusion}
In this work, we have shown the first high-contrast images of the $\gamma$ Cas analog RX J1744.7$-$2713, using observations from the Keck/NIRC2 vortex coronagraph. These images reveal the presence of 21 sources detected with a SNR $>$5. From a combination of various analyses, we found that three of these source are strong candidates as stellar companions, and that they would be in wide orbits ($\sim$ 450 AU to 2500 AU). Follow-up observations will be needed in $\sim$ 5--10 years to conduct a more significant proper motion analysis and confirm the nature of these detections. We also aim to obtain the spectrum of the sources in order to assess the nature of the sources. If confirmed, these results have direct implications for our understanding of how multiple star systems form and how X-ray binaries may actually not be simple binary systems. These observations are part of a pilot study aiming to provide direct imaging for all X-ray binaries within $\sim$ 3 kpc accessible with Keck/NIRC2. To date, 14 X-ray binaries have been observed, and the remaining results will be presented in a future paper (Prasow-Émond et al. in prep.). 

\acknowledgments
We want to thank V. Christiaens for his help with VIP.

M.P.E. is supported by the Institute for Data Valorisation (IVADO) through the M. Sc. Excellence Scholarship, by the Department of Physics of the Université de Montréal and by the Institute for Research on Exoplanets (iREx). J.H.L. is supported by the Natural Sciences and Engineering Research Council of Canada (NSERC) through the Canada Research Chair programs and wishes to acknowledge the support of an NSERC Discovery Grant and the NSERC accelerator grant. J.R. acknowledges support by the French National Research Agency in the framework of the Investissements d’Avenir program (ANR-15-IDEX-02), through the funding of the "Origin of Life" project of the Univ. Grenoble-Alpes. D.J.W. acknowledges financial support from STFC in the form of an Ernest Rutherford fellowship.

\bibliography{bib}
\bibliographystyle{aasjournal}
\appendix

\begin{figure}[h!]
    \centering
    \includegraphics[width = 0.49\textwidth]{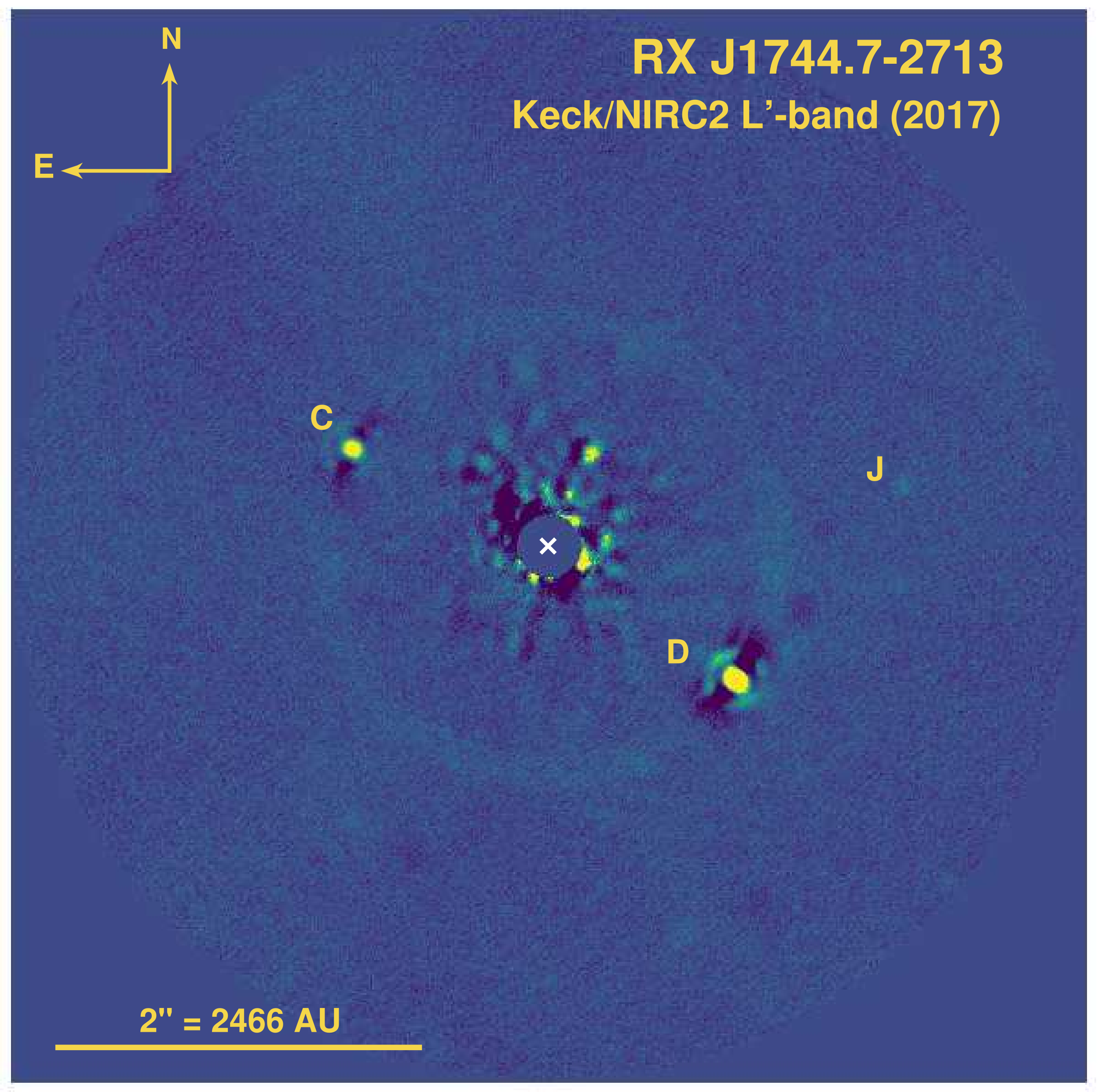}
    \caption{Keck/NIRC2 $L'$ high-contrast image of RX J1744.7$-$2713 acquired on September 08 2017 and treated and reduced with a PCA annular ADI algorithm \citep{2017AJ....154....7G}. Labeled are the sources detected with a SNR $>$ 5, using the same labels from Fig. \ref{fig:images}. The white X symbol denotes the position of the X-ray binary masked by the coronagraph.}
\end{figure}

    \begin{table*}
        \caption{Physical properties of the 21 sources in RX J1744.7$-$2713, where $r$ is the radial separation from the X-ray binary in mas, P.A. is the position angle in the image in degrees, and $L'$ and $Ks$ are the apparent magnitudes (see Section \ref{sec:injection} for errors). No extinction correction was applied (see Section \ref{sec:mag}). The second column is the optimal number of principal components used for the fitting.}
        \centering
        \begin{tabular}{cccccccc}
        \hline \hline
       Source & $n_\mathrm{comp}$ & $r$ & P.A. & $L'$ & $Ks$ & Estimated mass & Projected separation\\
       & & (mas) & (deg) & (mag) & (mag) & ($M_\mathrm{\odot}$) & (AU) \\ \hline
        B & 14 & 368 $\pm$ 4 & 340.5 $\pm$ 1.4 & 13.5 $\pm$ 0.7 & 13.2 $\pm$ 0.7 & 0.9--1.2 & 449 \\
        C & 4 & 1180 $\pm$ 6 & 154.2 $\pm$ 0.4 & 12.2 $\pm$ 0.6 & 12.2 $\pm$ 0.6 & 1.7--2.2 & 1440 \\
        D & 2 & 1246 $\pm$ 5 & 324.7 $\pm$ 0.5 & 10.7 $\pm$ 0.5 & 10.7 $\pm$ 0.5 & 4.1--4.8 & 1520 \\
        E & 5 & 3018 $\pm$ 2 & 140.6 $\pm$ 0.2 & 11.9 $\pm$ 0.6 & 11.8 $\pm$ 0.6 & 2.5--2.6 & 3681 \\
        F & 15 & 3189 $\pm$ 7 & 341.8 $\pm$ 0.2 & 12.2 $\pm$ 0.6 & 12.4 $\pm$ 0.6 & 1.8--2.2 & 3890 \\
        G & 29 & 4506 $\pm$ 12 & 359.1 $\pm$ 0.1 & 12.0 $\pm$ 0.6 & 12.4 $\pm$ 0.6 & 1.9--2.4 & 5497 \\
        H & 25 & 4228 $\pm$ 9 & 201.34 $\pm$ 0.09 & 15.5 $\pm$ 0.8 & 15.2 $\pm$ 0.8 & 0.2--0.6 & 5158 \\
        I & 16 & 3369 $\pm$ 4 & 148.4 $\pm$ 0.1 & 15.9 $\pm$ 0.8 & 15.6 $\pm$ 0.8 & 0.1--0.5 & 4110 \\
        J & 16 & 1956 $\pm$ 12 & 10.0 $\pm$ 0.1 & 14.8 $\pm$ 0.7 & 14.8 $\pm$ 0.7 & 0.3--0.8 & 2386 \\
        K & 28 & 3933 $\pm$ 10 & 291.9 $\pm$ 0.3 & 15.5 $\pm$ 0.8 & 15.4 $\pm$ 0.8 & 0.2--0.6 & 4798 \\
        L & 8 & 2851 $\pm$ 1 & 320.1 $\pm$ 0.2 & 16.0 $\pm$ 0.8 & 15.6 $\pm$ 0.8 & 0.1--0.5 & 3478 \\
        M & 33 & 4416 $\pm$ 5 & 350.3 $\pm$ 0.1 & 13.7 $\pm$ 0.7 & 14.0 $\pm$ 0.7 & 0.8--1.1 & 5387 \\
        N & 7 & 2803 $\pm$ 5 & 257.5 $\pm$ 0.2 & 15.8 $\pm$ 0.8 & 15.5 $\pm$ 0.8 & 0.2--0.5 & 3419 \\
        O & 25 & 4291 $\pm$ 8 & 287.0 $\pm$ 0.3 & 16.0 $\pm$ 0.8 & 15.6 $\pm$ 0.8 & 0.1--0.5 & 5235 \\
        P & 21 & 1835 $\pm$ 8 & 319.0 $\pm$ 0.2 & 15.8 $\pm$ 0.8 & 15.6 $\pm$ 0.8 & 0.2--0.5 & 2238 \\
        Q & 20 & 2176 $\pm$ 16 & 316.5 $\pm$ 0.6 & 15.5 $\pm$ 0.8 & 15.6 $\pm$ 0.8 & 0.2--0.6 & 2655 \\
        R & 26 & 1910 $\pm$ 17 & 354.18 $\pm$ 0.09 & 15.7 $\pm$ 0.8 & 15.6 $\pm$ 0.8 & 0.2--0.5 & 2331 \\
        S & 22 & 3737 $\pm$ 15 & 133.95 $\pm$ 0.05 & 16.0 $\pm$ 0.8 & 15.6 $\pm$ 0.8 & 0.1--0.5 & 4559 \\
        T & 32 & 4016 $\pm$ 9 & 149.5 $\pm$ 0.1 & 15.8 $\pm$ 0.8 & 15.5 $\pm$ 0.8 & 0.2--0.5 & 4899 \\
        U & 17 & 3689.3 $\pm$ 0.2 & 177.87 $\pm$ 0.01 & 15.7 $\pm$ 0.8 & 15.6 $\pm$ 0.8 & 0.2--0.5 & 4501 \\
        V & 8 & 2475 $\pm$ 18 & 186.6 $\pm$ 0.1 & 15.5 $\pm$ 0.8 & 15.6 $\pm$ 0.8 & 0.2--0.6 & 3020 \\ \hline
        \end{tabular}
       \label{tab:candidates}
   \end{table*}
   
\newpage
   
\begin{figure*}
    \centering
    \includegraphics[width=0.32\textwidth]{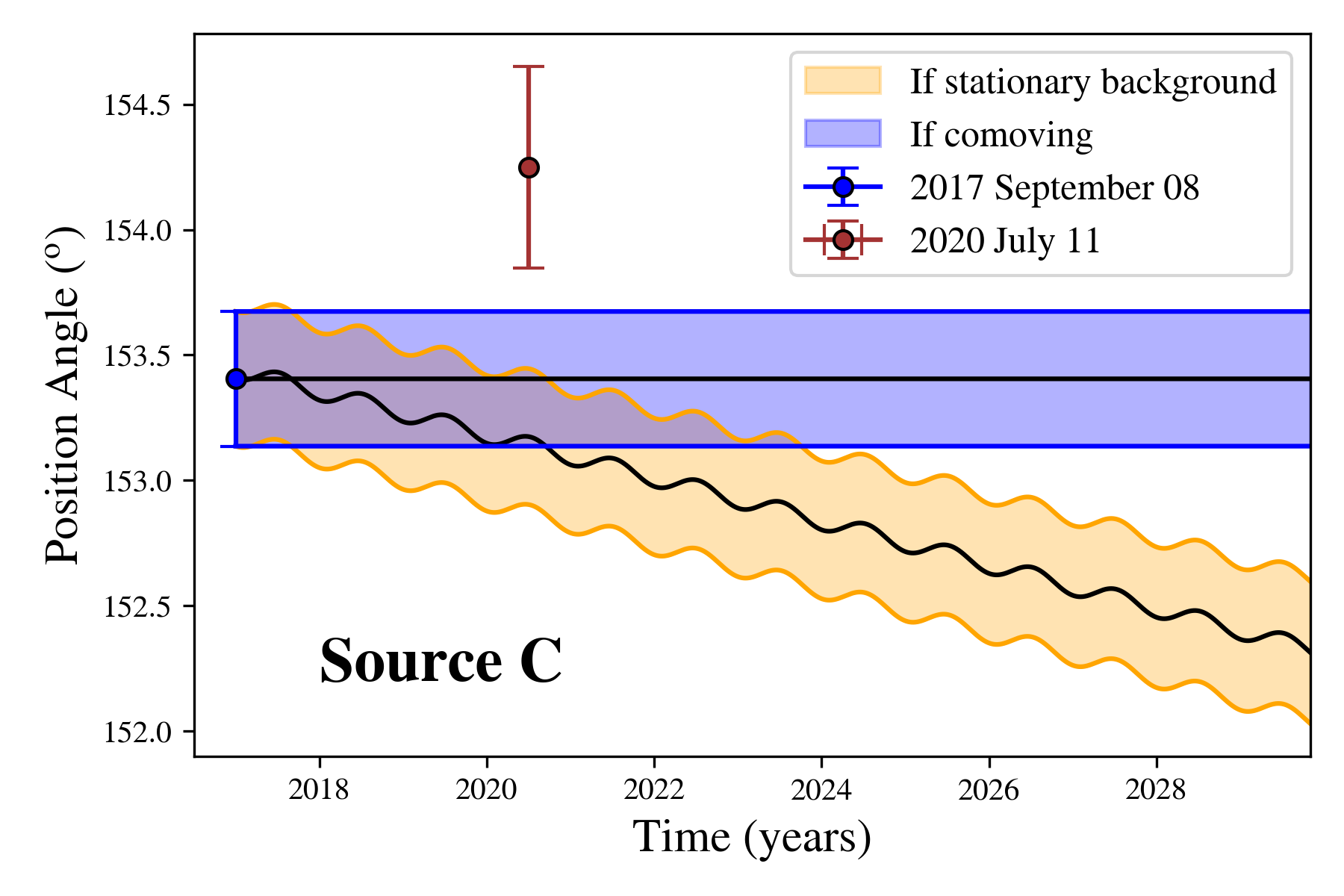}
    \includegraphics[width=0.32\textwidth]{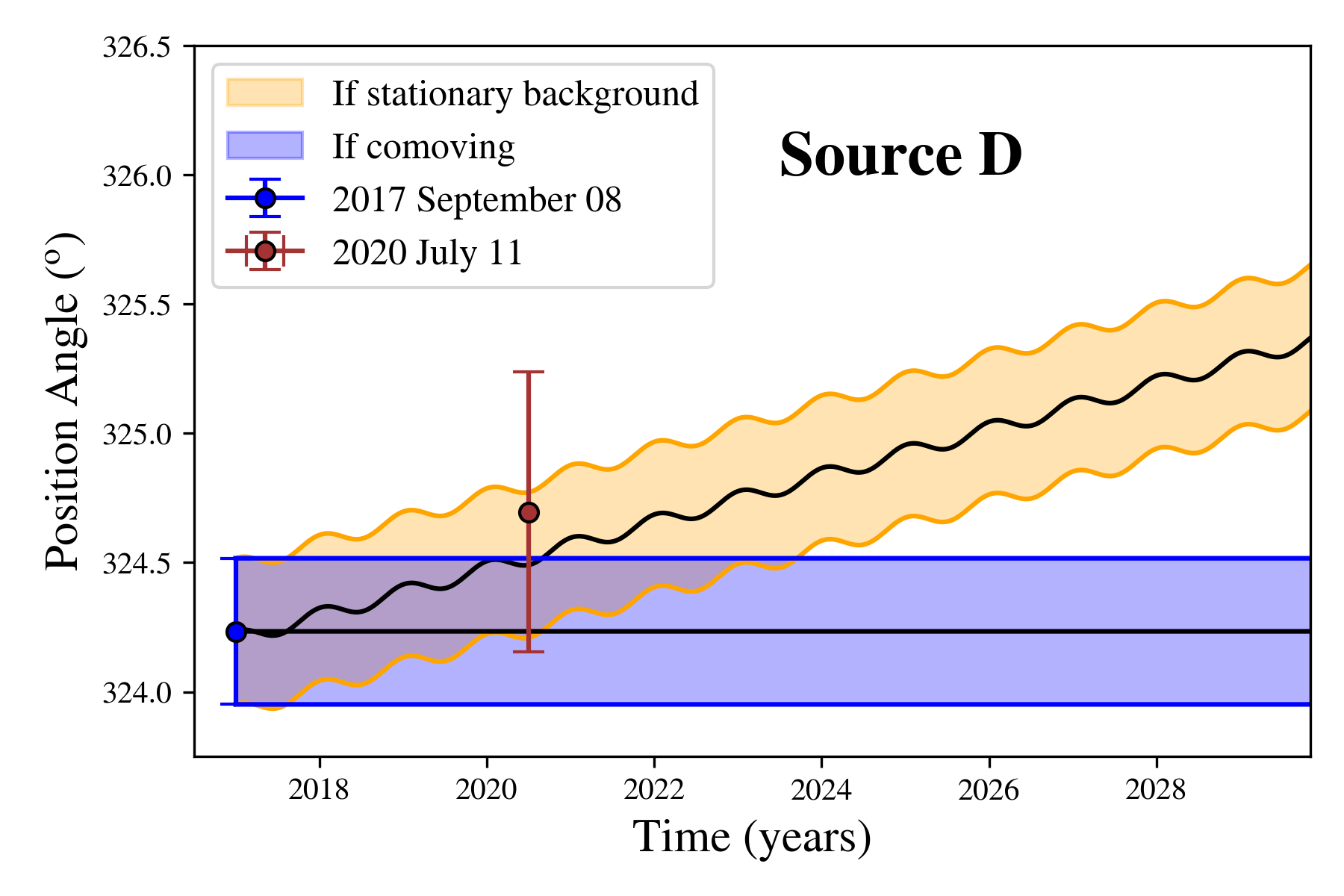}
    \includegraphics[width=0.32\textwidth]{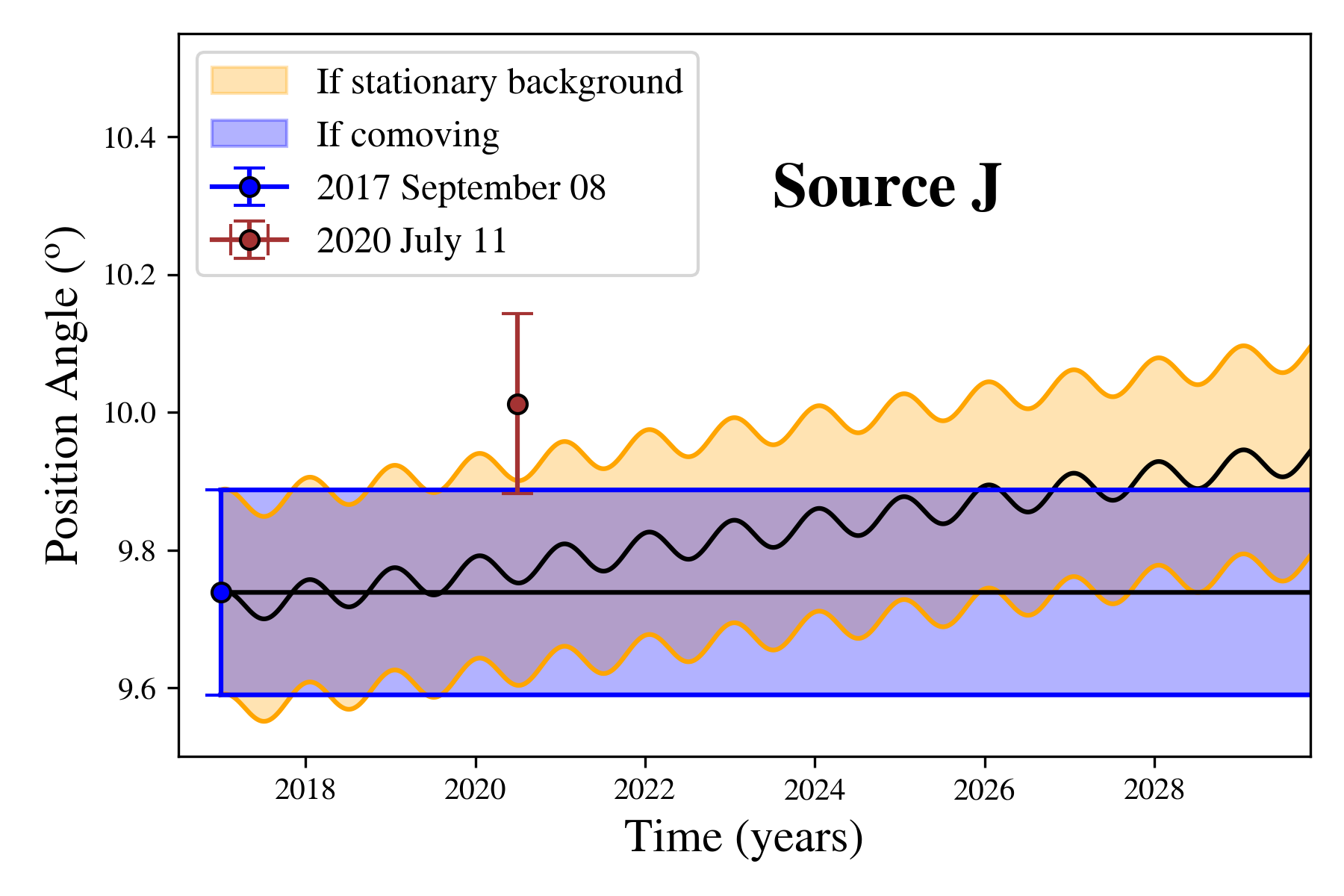}
    \includegraphics[width=0.32\textwidth]{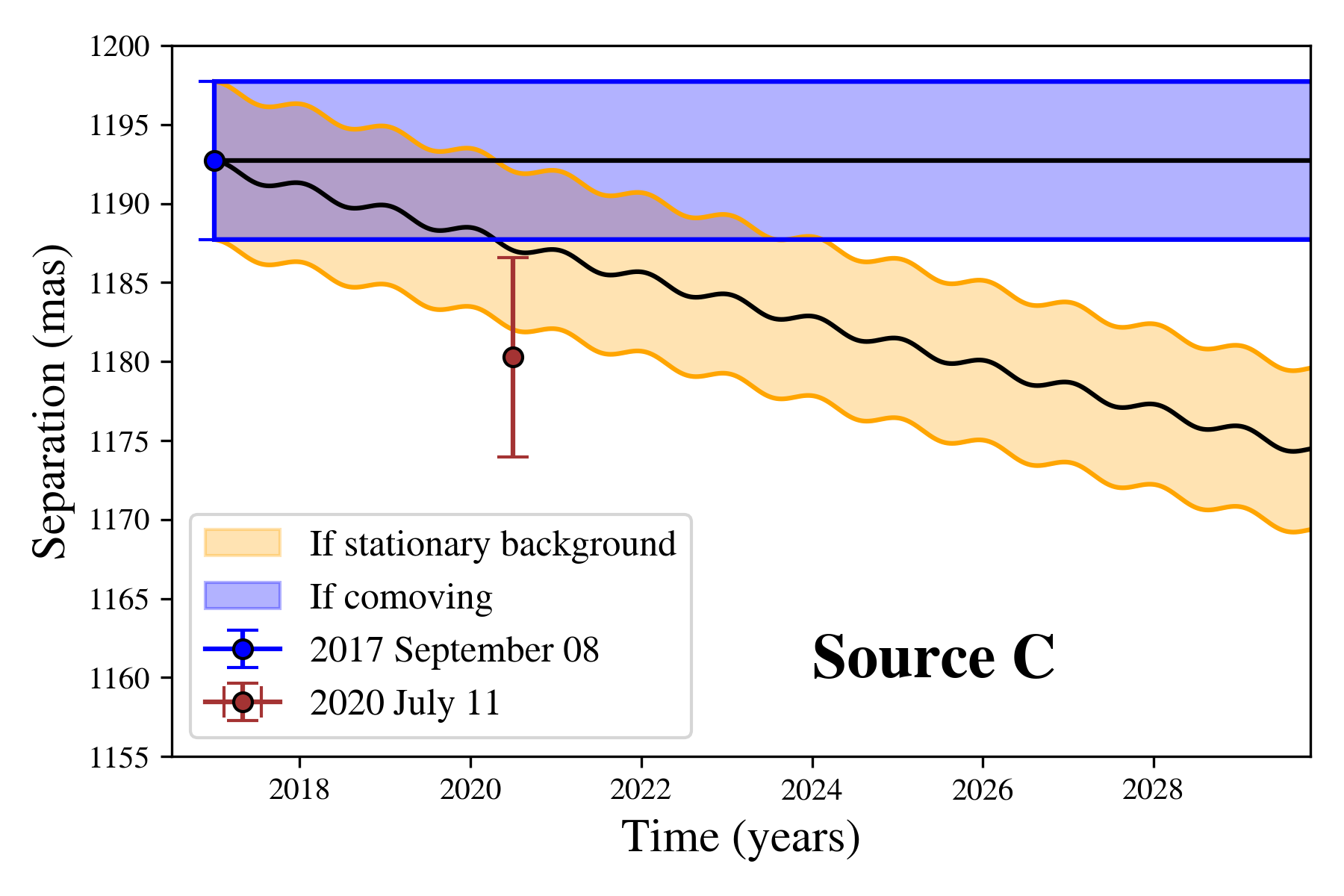}
    \includegraphics[width=0.32\textwidth]{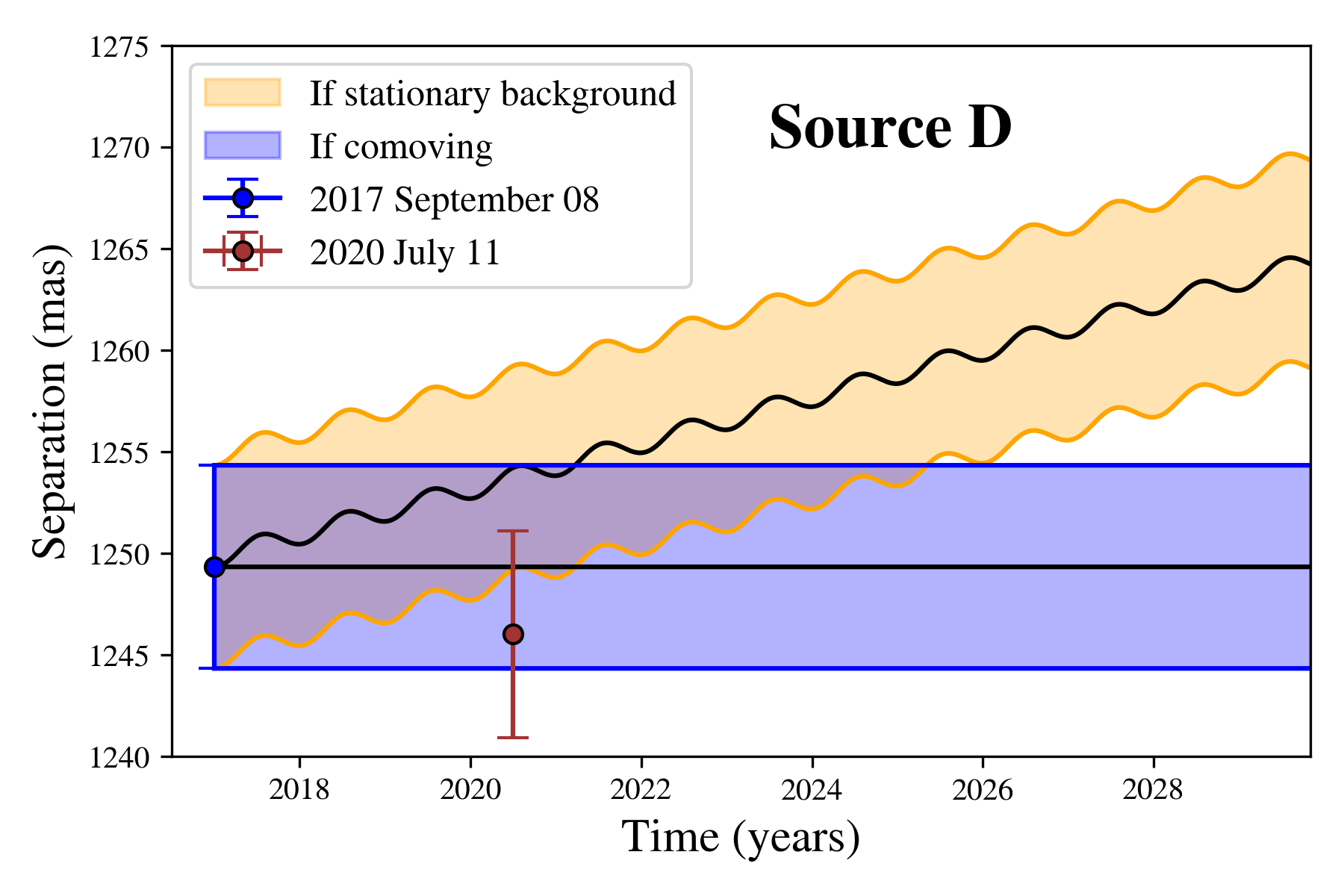}
    \includegraphics[width=0.32\textwidth]{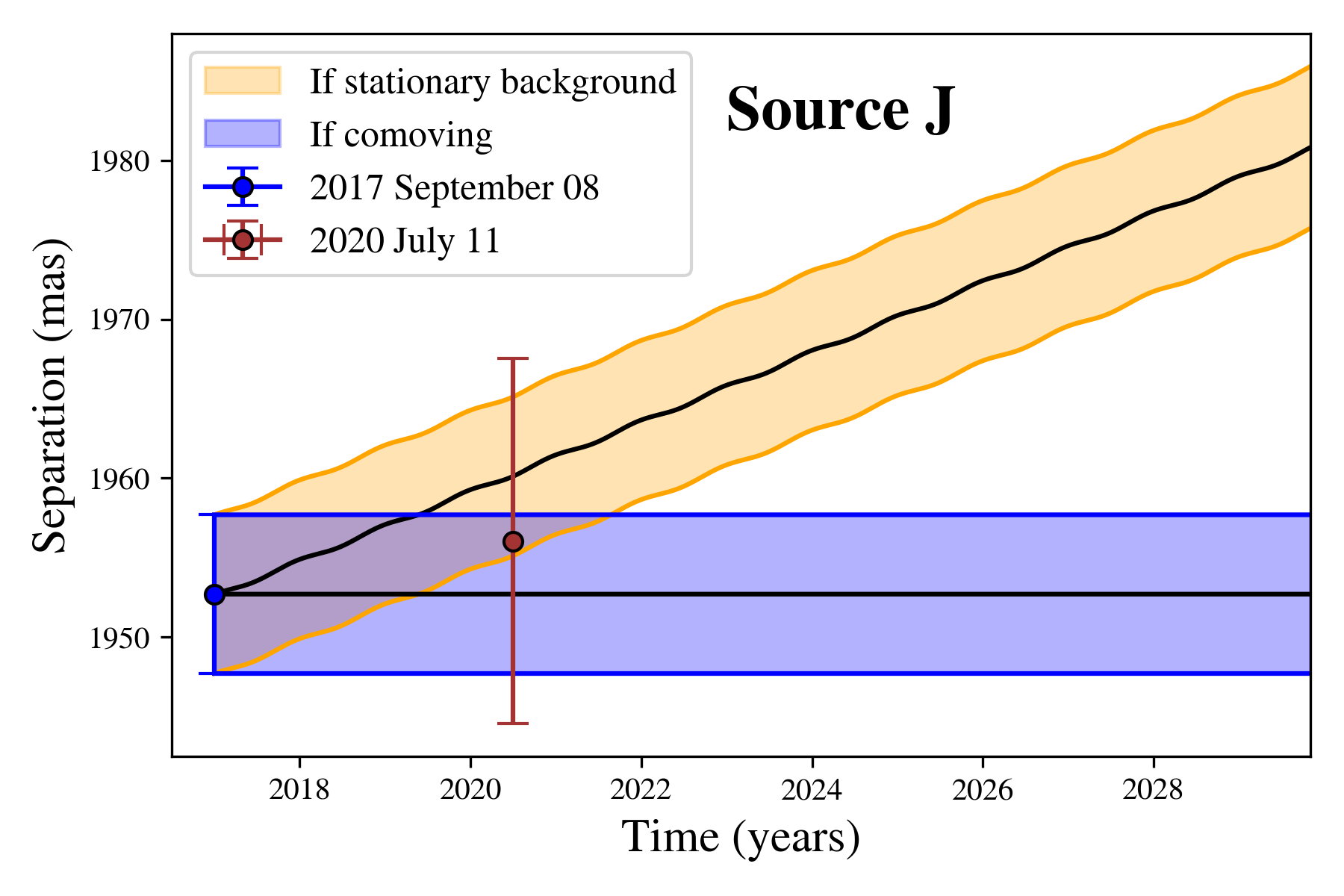}
    \includegraphics[width=0.32\textwidth]{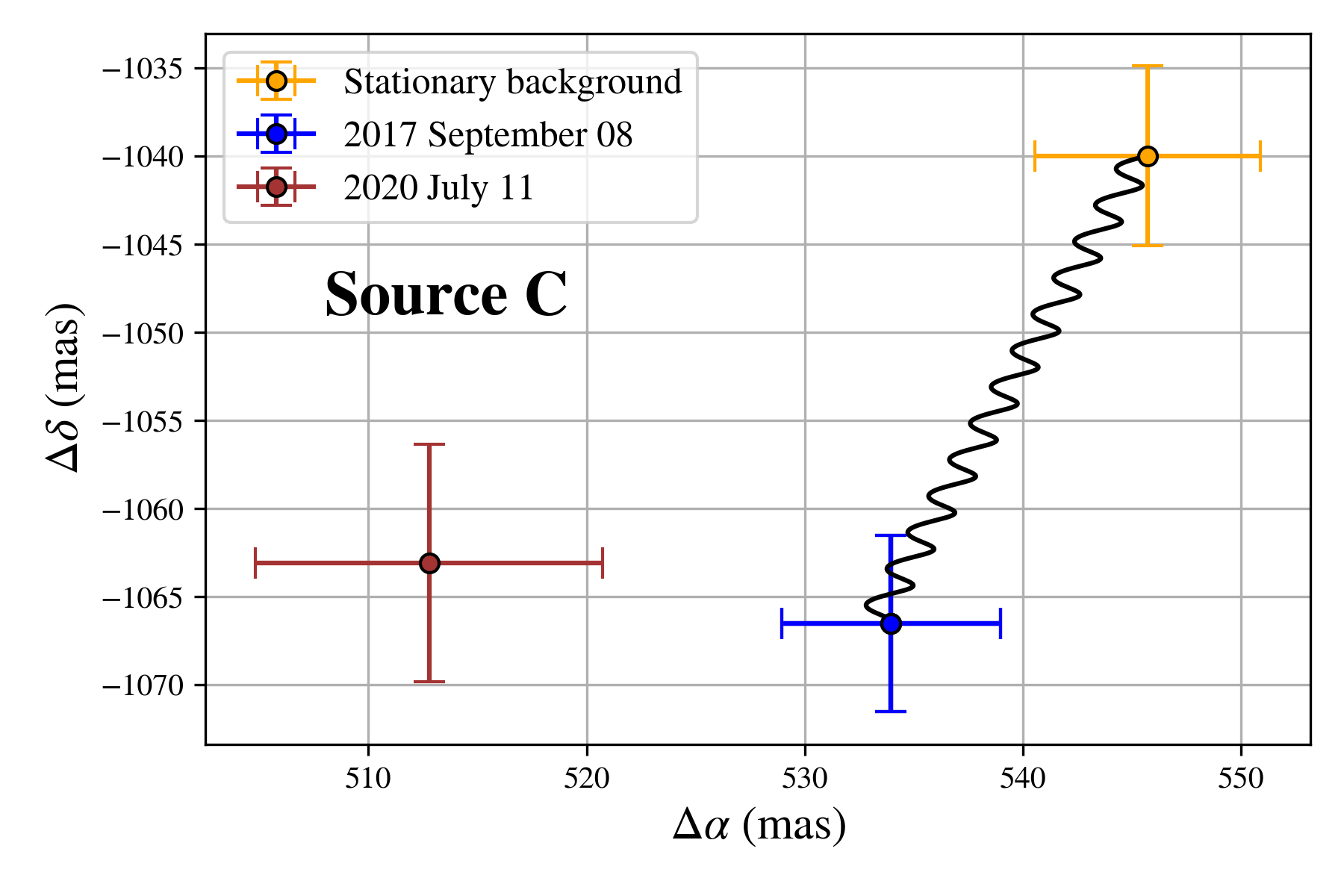}
    \includegraphics[width=0.32\textwidth]{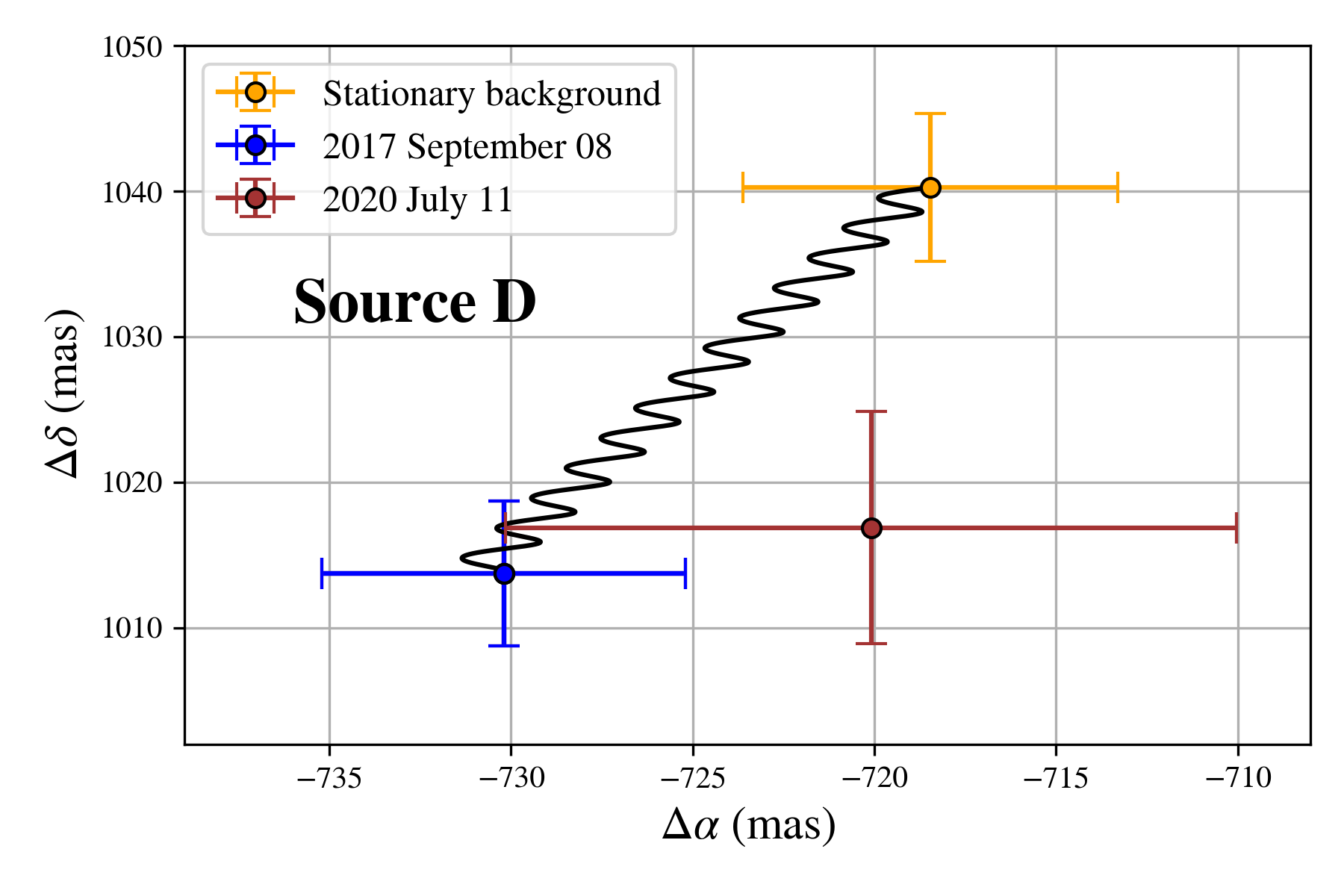}
    \includegraphics[width=0.32\textwidth]{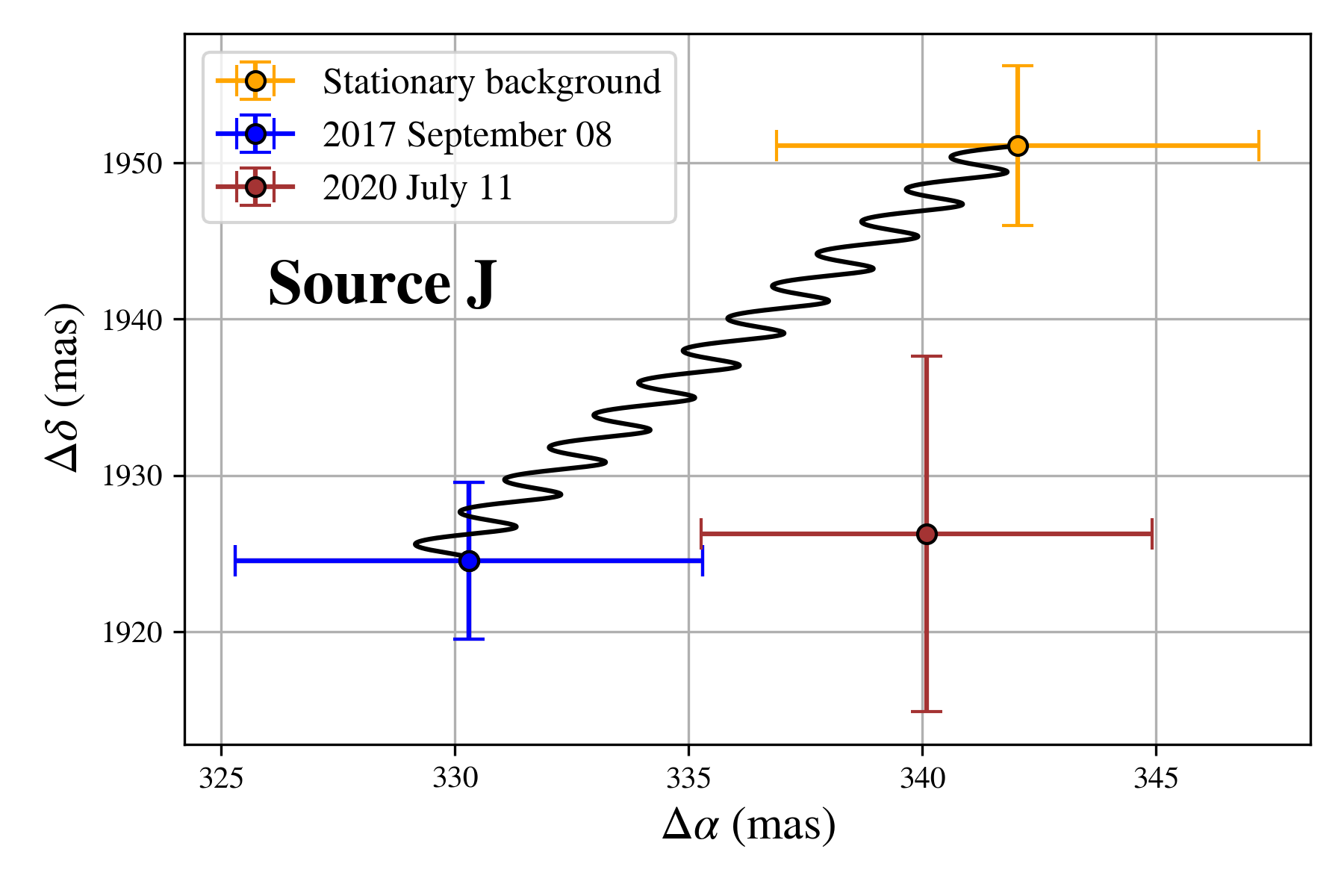}
    \caption{\textit{Top row}: Position angle in degrees of the sources C (left), D (middle), and J (right) relative to its host, as measured by Keck/NIRC2 on 2017 September 08 (blue data point). The red data point is the position angle measured by Keck/NIRC2 on 2020 July 11. An infinitely distant background object with zero proper motion would be following the orange track, while a comoving companion would lie within the blue area. \textit{Middle row}: Same as the top row, but for the separation from the X-ray binary in mas. \textit{Bottom row}: Relative separations between the source and the X-ray binary in right ascension ($\alpha$) and declination ($\delta$). The first epoch astrometric point is plotted in blue (2017 September 08), while the second epoch astrometric point is plotted in red (2020 July 11). The excepted position for a background object is plotted in yellow, along with a proper and parallactic motion track.}
    \label{fig:TRILEGAL}
\end{figure*}

\end{document}